\documentclass[reprint, prl, superscriptaddress, floatfix]{revtex4-1}

\usepackage{hyperref}
\usepackage{graphicx}
\usepackage{amsmath}
\usepackage{wasysym}
\usepackage{amsfonts}
\usepackage{color}
\usepackage{verbatim}
\usepackage{upgreek}
\usepackage{bm}

\newcommand{\beq}{\begin{equation}}
\newcommand{\eeq}{\end{equation}}
\newcommand{\bea}{\begin{eqnarray}}
\newcommand{\eea}{\end{eqnarray}}
\newcommand{\bal}{\begin{align}}
\newcommand{\eal}{\end{align}}

\newcommand{\fig}[1]{Fig.~\ref{#1}}

\newcommand{\WSe}{WSe$_2$}

\begin{document}
\title{Momentum-Resolved View of Electron-Phonon Coupling in Multilayer \WSe}
\author{L. Waldecker} 
\email{waldecker@stanford.edu}
\affiliation{Fritz Haber Institut of the Max Planck Society, Faradayweg 4-6, 14195 Berlin, Germany}
\affiliation{Stanford University,  348 Via Pueblo Mall, Stanford, California 94305, USA}
\author{R. Bertoni}
\affiliation{Fritz Haber Institut of the Max Planck Society, Faradayweg 4-6, 14195 Berlin, Germany}
\affiliation{Univ Rennes 1, CNRS, Institut de Physique de Rennes, UMR 6251, UBL, F-35042 Rennes, France}
\author{H. H\"ubener}
\affiliation{Max Planck Institute for the Structure and Dynamics of Matter and Center for Free-Electron Laser Science, Notkestraße 85, 22761 Hamburg, Germany}
\author{T. Brumme}
\affiliation{Max Planck Institute for the Structure and Dynamics of Matter and Center for Free-Electron Laser Science, Notkestraße 85, 22761 Hamburg, Germany}
\author{T. Vasileiadis}
\affiliation{Fritz Haber Institut of the Max Planck Society, Faradayweg 4-6, 14195 Berlin, Germany}
\author{D. Zahn}
\affiliation{Fritz Haber Institut of the Max Planck Society, Faradayweg 4-6, 14195 Berlin, Germany}
\author{A. Rubio}
\affiliation{Max Planck Institute for the Structure and Dynamics of Matter and Center for Free-Electron Laser Science, Notkestraße 85, 22761 Hamburg, Germany}
\author{R. Ernstorfer}
\email{ernstorfer@fhi-berlin.mpg.de}
\affiliation{Fritz Haber Institut of the Max Planck Society, Faradayweg 4-6, 14195 Berlin, Germany}

\begin{abstract}

We investigate the interactions of photoexcited carriers with lattice vibrations in thin films of the layered transition metal dichalcogenide (TMDC) \WSe. Employing femtosecond electron diffraction with monocrystalline samples and first principle density functional theory calculations, we obtain a momentum-resolved picture of the energy-transfer from excited electrons to phonons. 
The measured momentum-dependent phonon population dynamics are compared to first principle calculations of the phonon linewidth and can be rationalized in terms of electronic phase-space arguments.
The relaxation of excited states in the conduction band is dominated by intervalley scattering between $\Sigma$ valleys and the emission of zone-boundary phonons.
Transiently, the momentum-dependent electron-phonon coupling leads to a non-thermal phonon distribution, which, on longer timescales, relaxes to a thermal distribution via electron-phonon and phonon-phonon collisions. Our results constitute a basis for monitoring and predicting out of equilibrium electrical and thermal transport properties for nanoscale applications of TMDCs.

\end{abstract}

\maketitle
\date{\today}
Semiconducting transition metal dichalcogenides (scTMDCs) combine crystal structures of chemically stable two-dimensional (2D) layers with indirect bandgaps in the visible and near infrared optical range. Their intrinsic stability down to monolayer thicknesses \cite{Mak2010, Splendiani2010} in combination with the possibility to create artificial stacks \cite{Geim2013, Britnell2013} suggests them for electronic and opto-electronic applications like nanoscale transistors or photodetectors with atomically sharp p-n junctions \cite{Radisavljevic2011, Koppens2014, Lee2014}. 
In such devices, electronic mobilities, electronic coupling and heat conductivities within the layers and across interfaces are of central interest. 
Whereas macroscopic heat- and electrical transport properties can be measured directly, the observation of the underlying microscopic processes, i.e., the scattering processes of carriers and of phonons, require methods with time-, momentum and energy-resolution to be understood in detail. 
Such information can be decisive in the correct determination of transport properties \cite{Piscanec2004, Huewe2015} or energy relaxation \cite{Sentef2013, Waldecker2016, Waldecker2017}.
A momentum-resolved view on scattering processes will in addition be of uttermost importance in conceiving novel quantum technologies harnessing spin- and valley-degrees of freedom \cite{Mai2014, Han2014, Ye2016}, as they utilize carrier populations at specific positions in momentum space. 
While time- and angle-resolved photoemission spectroscopy provides this level of detail for electron dynamics, see for instance \cite{Schmitt2008,Bertoni2016ARPES}, techniques for studying ultrafast structural dynamics have not yet reached the equivalent level of resolution. Recently, the investigation of time- and momentum-resolved phonon population has been demonstrated with ultrafast x-ray and electron diffraction \cite{Trigo2010, Zhu2015, Chase2016, Harb2016}.

This work reports a momentum-resolved study of scattering processes and the resulting energy-transfer between photoexcited electrons and phonons in thin bulk-like films of \WSe. Specifically, we probe the in-plane structural dynamics following resonant electronic excitation with a short (50~fs) laser pulse by femtosecond electron diffraction (FED), described in \cite{Waldecker2015Setup}. 
The spectrum of the excitation pulses is centered at a photon energy of $1.55$~eV, with significant spectral weight at the A-exciton transition at 1.59~eV \cite{Frindt1963}. 
The incident pump fluence applied in this study was 7~mJ/cm$^2$, which is most likely high enough to significantly lower the exciton binding energy or even lead to exciton dissociation \cite{Chernikov2015}. 
The linearly polarized pump pulses create electronic excitations in the K and K' valleys of the conduction band. 
Time- and angle-resolved photoemission spectroscopy studies have shown that carriers scatter from the K valleys to the global minimum of the conduction band (the $\Sigma$ valleys) within less than 100~fs \cite{Bertoni2016ARPES, Wallauer2016}.  
Electron relaxation in multilayer \WSe\ at times larger than 100~fs is thus dominated by phonon-mediated relaxation of electrons to the bottom of the conduction band within the $\Sigma$ valleys. 

\begin{figure}[bth]
\begin{center}
\includegraphics[width=1.0\columnwidth]{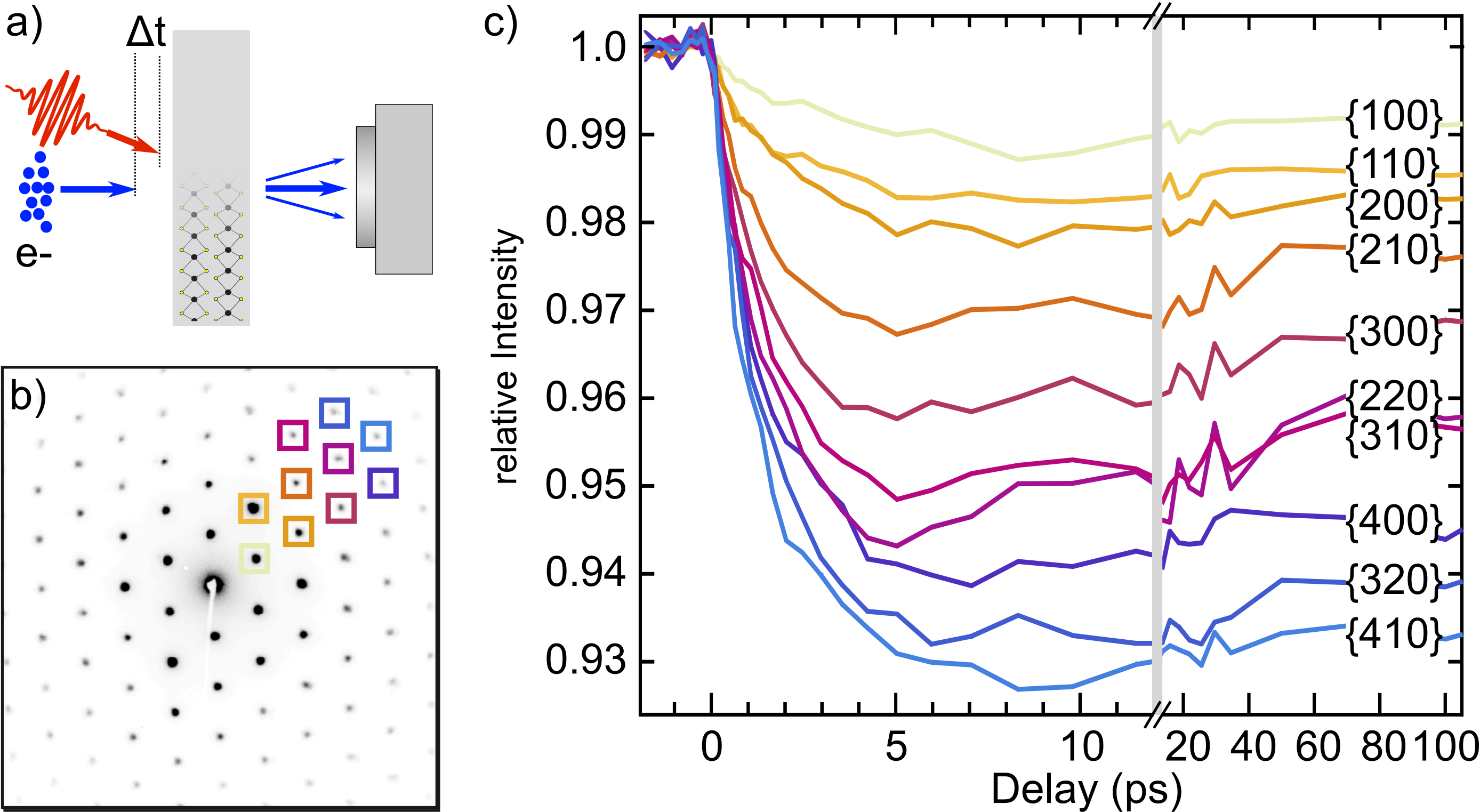}
\caption{a) Schematical representation of the experiment. b) Diffraction image of thin \WSe\ films. c) Evolution of the averaged intensity of different families of Bragg peaks, as indicated by the colored boxes in b). }
\label{fig:pdecay}
\end{center}
\end{figure}

\begin{figure}[bth]
\begin{center}
\includegraphics[width=1.0\columnwidth]{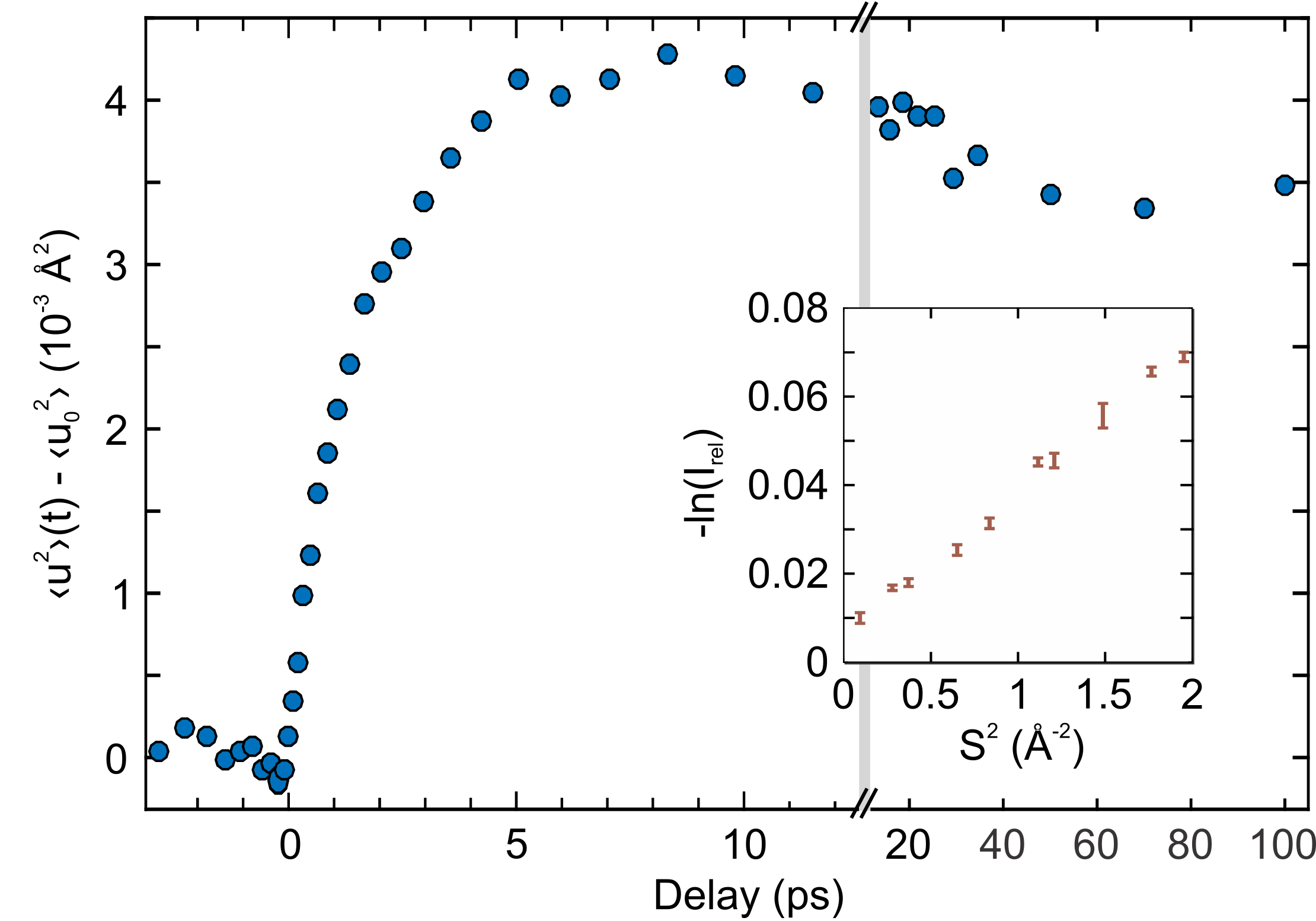}
\caption{Temporal evolution of the effective atomic mean square displacement. The inset shows the intensity of different families of Bragg peaks as a function of squared scattering vector for pump-probe delays larger than 20 ps. All peaks are used to calculate the MSD.}
\label{fig:msd}
\end{center}
\end{figure}

The samples were produced by mechanical exfoliation from bulk crystals (HQ Graphene) down to thicknesses of around 30 nm, estimated by their optical transmission. Samples of sizes larger than 150x150 $\upmu$m$^2$ were identified and transferred onto standard copper grids. Diffraction images were taken at normal incidence of the electron pulses to the sample at various delay times with respect to the optical excitation, see \fig{fig:pdecay}~a. 
The energy of the electrons was 95~keV, resulting in a nearly flat Ewald sphere in reciprocal space and thus a large number of visible Bragg peaks, sensitive to in-plane dynamics. 
From the raw images (\fig{fig:pdecay}~b) the evolution of the Bragg peak intensities are extracted by integrating the images in circular areas around each peak and subsequent averaging of all peaks of the same family (see Figure \ref{fig:pdecay}). 
The negative logarithm of the relative intensity at large delay times $-\ln(I_{rel}(t \geq 20 \mathrm{ps}))/s^2$, scaled with the scattering vector of the respective family of peaks, is shown in the inset of \fig{fig:msd}. 
Its linear increase indicates that the experiment can be described in the single-scattering limit.  
A similar behavior of all peaks within each family (not shown) points to an isotropic increase of lattice vibrations in the measurement plane. 
The evolution of the Bragg peak intensities is used to calculate the change of the effective atomic mean square displacement (MSD) $ \langle u^2 \rangle(t) - \langle u_{RT}^2 \rangle = -3/(4 \pi^2)\cdot \ln(I_{rel}(t))/s^2$. 
The result is shown as a function of delay time in Figure \ref{fig:msd}. 
The rise of MSD proceeds within several picoseconds and is related to the energy transfer from excited electrons to phonons. 
A maximum is reached between 5 and 10~ps, after which the MSD slightly decreases to reach a new steady-state at around 50 ps. 
Heat diffusion out of the free-standing films is not expected to play an important role on the timescales relevant for the measurements here.
A fit of an exponential rise and decay to the MSD yields time constants of $\tau_1 = 1.83 \pm 0.13$~ps and $\tau_2 = 19 \pm 5$~ps. 
As electrons scatter to the $\Sigma$ valleys of the conduction band within less than 100~fs \cite{Bertoni2016ARPES}, but no significant changes of the MSD are observed for these delay times, we conclude that the energy transfer from electrons to phonons is dominated by electrons scattering from initial states in the $\Sigma$ valleys to energetically available final states.

\begin{figure*}[bth]
\begin{center}
\includegraphics[width=1.0\textwidth]{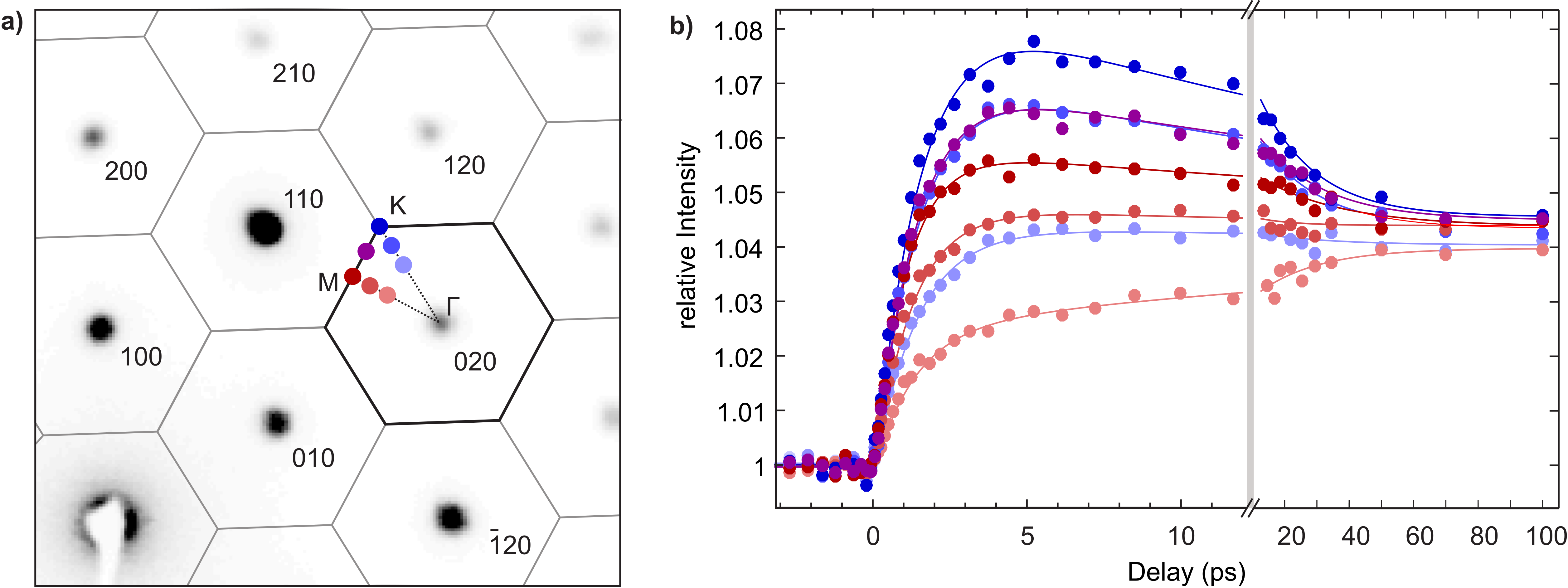}
\caption{a) Detail of a diffraction pattern of \WSe\ with the Brillouin zone boundaries overlayed. Points of analysis along the high-symmetry lines are marked with colorored circles. b) Evolution of the scattered intensity (circles) with pump-probe delay at different positions of the Brillouin zone (same color code as in panel a)\,). The solid lines are bi-exponential fits to the data.}
\label{fig:bg}
\end{center}
\end{figure*}

The intensity in a diffraction image can be decomposed into a first order process - diffraction into the Bragg peaks, satisfying the Laue condition - and second order processes including inelastic scattering with phonons or with short range correlations, such as crystal defects. These secondary processes result in scattered intensity between Bragg peaks. The conservation of total momentum requires a probe-electron scattering with a phonon to obey the modified Laue condition $\bm k_{pr} - \bm k'_{pr}=\bm G_{hkl} + \bm q$, where $\bm k_{pr}$ and $\bm k'_{pr}$ are the wave vectors of the electrons of the probe pulse, $\bm G_{hkl}$ is a reciprocal lattice vector and $\bm q$ is the wave vector of a phonon \cite{Trigo2010}. The intensity in between Bragg-peaks therefore contains information on the population of phonons as a function of their wavevector  \cite{Xu2005, Trigo2010, Zhu2015, Chase2016, Harb2016}.
In figure~\ref{fig:bg}~a, a detail of a diffraction pattern of \WSe\ is shown. 
By integrating circular areas around points along the high-symmetry directions $\Gamma$-K, $\Gamma$-M and K-M and subsequently averaging all symmetry-equivalent areas, the phonon dynamics in the irreducible part of the BZ are obtained. The limited coherence of the source does not allow for the investigation of points close to the $\Gamma$ point, as these are dominated by the intensity of the tail of the Bragg-peaks. 
The obtained relative intensities are plotted in Figure \ref{fig:bg}~b as  a function of delay time. 
Their evolution shows a clear ${\mathbf q}$-point dependence, with the main increase in the first 10 ps being observed close to the border of the BZ. 
On longer delays (10-50 ps), intensity is shifted away from the border of the BZ and a steady state is reached at approximately 50 ps, indicating that the lattice has reached an apparent thermal state at an elevated temperature.

\begin{figure*}[bth]
\begin{center}
\includegraphics[width=1.0\textwidth]{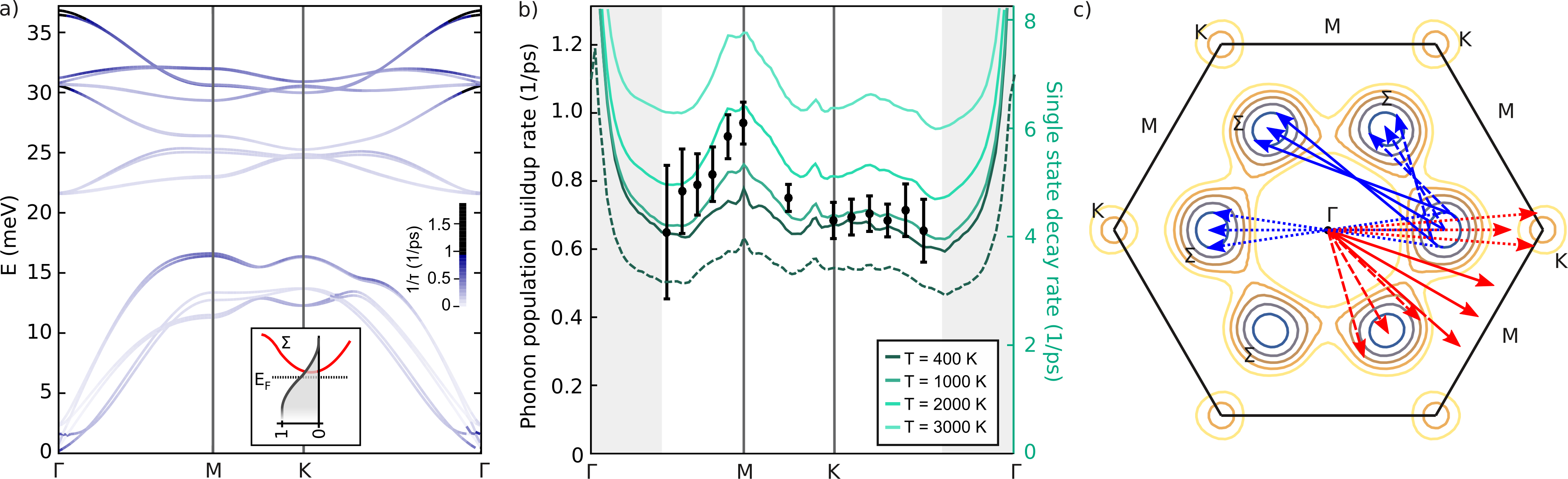}
\caption{a) Calculated phonon bandstructure along the high-symmetry directions. The colour code shows the calculated scattering rates, Eq.~(1), of each phonon mode caused by scattering with excited electrons in the conduction band for an electronic temperature of 1000K (see sketch of the modeling in the inset). b) Experimentally determined values of the momentum-dependent rate of population build-up (black circles with error bars) and first principles calculation of the single-state phonon scattering rates for different electronic temperatures. 
Solid lines are summed over all phonon branches and the dashed line is the sum of all branches, weighted with their phonon polarization vector. The gray areas indicate the \textbf{q}-range not accessible in the experiment. c) Constant energy contours of the lowest conduction band showing the K and $\Sigma$ valleys. 
Exemplary electronically allowed scattering processes of electrons situated in a $\Sigma$ valley are depicted by blue arrows and the phonon momenta required to obey conservation of momentum are shown as red arrows. }
\label{fig:tconst}
\end{center}
\end{figure*}

To quantify the coupling between electrons and phonons at different ${\mathbf q}$-points, we fit the curves shown in Figure \ref{fig:bg}~b with two exponential functions $a_{{\mathbf q},i} \exp(t/\tau_{{\mathbf q},i})$, which, to first approximation, represent the effects of (1) electron-phonon coupling to phonons of wavevector $\mathbf q$ and (2) phonon-phonon relaxation. 
Figure \ref{fig:tconst} b shows the experimentally retrieved rates ($1/\tau_{{\mathbf q},1}$) of the momentum-dependent rise of the phonon population.
We note that these experimentally retrieved rates are the result of the averaged energy transfer from excited electrons to all phonon branches $j$ at the respective position in momentum space, which contribute to the measured signal by $I({\mathbf Q}) \propto \sum_j \frac{1}{\omega_j({\mathbf q})} \left[ n_j({\mathbf q} + \frac12) \right] |F_j({\mathbf Q})|^2$ \cite{Trigo2010}, where $\omega_j$ and $n_j$ are the frequency and occupation number of the phonon mode with wavevector ${\mathbf q}$, ${\mathbf Q} = {\mathbf q + \mathbf G_{hkl}}$, and $F_j({\mathbf Q})$ depends on the scalar product of the phonon polarization vector and $\mathbf Q $\footnote{$F_j(\mathbf{Q}) = \sum_s \frac{f_s}{\sqrt{m_s}} e^{-M_s}\left({\mathbf Q}\cdot \hat{\mathbf{e}}_{s,j,\mathbf{q}}\right)e^{-i \mathbf{G}\cdot \mathbf{r}_s}$ where $f_s$, $m_s$, $M_s$ and $\mathbf{r}_s$ denote the atomic scattering factor, the mass, the Debye-Waller factor and the position of atom $s$, $\hat{\mathbf{e}}_{s,j,\mathbf{q}}$ is the phonon polarization vector, and $\mathbf{G}$ is the closest reciprocal-lattice vector to $\mathbf{Q}$, i.e. $\mathbf{Q}=\mathbf{G}+\mathbf{q}$}. 
The $1/\omega_j({\mathbf q})$-dependence of the signal implies that the experiment is more sensitive to acoustic phonon modes. Furthermore, in thermal equilibrium, acoustic phonons have higher occupation than optical phonons and consequently contribute more to the scattered intensity.
To resolve the coupling to different phonon branches at a given wavevector $\mathbf q$ and the decay of optical into acoustic phonon modes, additional energy-resolution would be required as obtained in state-of-the-art transmission electron microscopy \cite{Krivanek2014}. 

We theoretically obtain the momentum- and state-resolved phonon lifetime $\tau_{\mathbf{q}\nu}^{\rm ph}$ from density functional theory (DFT) by computing
\bea
\label{eq:eph}
\frac{1}{\tau_{\mathbf{q}\nu}^{\rm ph}} && = \frac{2\pi}{\hbar}2\Sigma_{mn}\int \frac{d\mathbf{k}}{\Omega_{\rm BZ}}|g_{mn\nu}(\mathbf{k},\mathbf{q})|^2(f_{n\mathbf{k}}-f_{m\mathbf{k}+\mathbf{q}}) \nonumber\\
&& \times \delta(\epsilon_{m\mathbf{k}+\mathbf{q}}-\epsilon_{n\mathbf{k}}-\hbar \omega_{\mathbf{q}\nu}),
\eea
where $g$ are the state and momentum-resolved electron-phonon matrix elements, $\epsilon$ the electronic bands, $\omega$ the energy of phonon mode $\nu$ and $f_{nk}=f(e_{nk},T)$ the Fermi-Dirac distributions describing the occupations of the electronic bands.
To achieve a well defined excited state occupation close to the experimental conditions, we describe the electronic distribution by a Fermi-Dirac function with the Fermi-level fixed 30~meV below the conduction band minimum (see inset of \fig{fig:tconst} a), leading to a weak occupation of the conduction bands while retaining full occupation of the valence bands. 
We use the EPW\cite{Giustino2007,Ponce2016} code of the QUANTUM ESPRESSO\cite{Gianozzi2009} package to evaluate Eq.~(1) on a fine mesh in the BZ. As the theory considers adiabatic, harmonic phonons \cite{Giustino2017}, phonon-phonon coupling, which leads to the relaxation to a thermal state, is not captured.
The electronic ground state is evaluated accounting for spin-orbit effects on a 12$\times$12$\times$4 mesh using the norm conserving fully relativistic pseudo potentials with an energy cutoff of 160~Ry (2177~eV) and the experimental lattice parameters $a=3.282$ \AA\ and $c=12.980$ \AA\cite{Sharma1999}. 
The phononic structure is calculated for  6$\times$6$\times$2 points in the BZ. The integral in Eq.~(1) is sampled with  80$\times$80$\times$60 interpolated $\mathbf{k}$-points using 44 Wannier functions, accounting for $d$-orbitals of the tungsten atoms and $p$-orbitals of the selenium.

The resulting scattering rates $1/\tau_{\mathbf{q}\nu}^{\rm ph}$ are shown in \fig{fig:tconst}~a for each mode along the high-symmetry directions in the Brillouion zone. 
The total momentum-resolved scattering rate is obtained by summing these scattering rates over all phonon branches  at each $\mathbf{q}$-point and the total rates for four different electronic temperatures are shown as solid lines together with the experimental data in \fig{fig:tconst}~b.  
The dashed line shows the total scattering rate as a weighted sum with the in-plane projection of the phonon polarization vector of each contributing phonon branch as measured in the experiment. We find that the weighting with the polarization vector does not significantly change the shape of the calculated lifetime, nor does weighting the branches with $1/\omega_j(\mathbf{q})$ (not shown), which suggests that our experiment probes a representative set of in-plane phonons. With increasing electron temperature, the coupling strength increases relatively homogeneously in the BZ. 
A good qualitative agreement of the experimental and calculated couplings is evident, with a plateau-like region on the $\Gamma$-K line and a broad peak around M. It can be understood by examining the structure of the conduction band of \WSe\ and the energetically allowed electronic scattering processes. 
An energy contour plot of the lowest conduction band, as calculated by DFT, is shown in \fig{fig:tconst} c. 
Subsequent to vertical excitation of excitons in the K valleys, excited states swiftly scatter to the $\Sigma$ valleys~\cite{Bertoni2016ARPES}. During the subsequent electron relaxation, electrons populate almost exclusively the $\Sigma$ valleys around the bottom of the conduction band, which is also the lowest energetic state for excitons in bulk \WSe\ \cite{Schuster2016}. 
Energy relaxation of the excited electrons therefore occurs through intravalley scattering and intervalley scattering between different $\Sigma$ valleys (shown as blue arrows). 
The change of electronic momentum by scattering between these valleys is absorbed by the emission (or absorption) of a phonon (shown as red arrows). 
We can thus phenomenologically identify areas of the BZ, where phonon-emission is predominately expected. These are rather broad areas around the M, K, and $\Sigma$ points as well as around $\Gamma$ (intravalley scattering, not shown). 
The spin-texture with non-equivalent $\Sigma$ and $\Sigma$'-valleys \cite{Bertoni2016ARPES} should furthermore favor intravalley scattering as well as scattering between second-next valleys, corresponding to phonon emission around M and $\Gamma$. 
Indeed, theory and experiment show a slightly higher scattering rate around M compared to K and $\Sigma$. 
Furthermore, our calculations predict the highest coupling rate to be around $\Gamma$, which is the region not accessible by our experiment.
Theoretically, however, the increased rates for small phonon momenta are readily understood as scattering within the same $\Sigma$ valleys according to Eq.~(\ref{eq:eph}).

The comparison of the measured phonon population dynamics and the calculated phonon lifetimes (\fig{fig:tconst}) shows a very similar momentum-dependence of both quantities. Quantitatively, the calculated phonon lifetimes exceed the population dynamics by a factor six to ten, depending on the electron temperature.  
In the notion of Bloch-Boltzmann-Peierls equations, electrons relax to the bottom of the conduction band by multiple subsequent scattering events with phonons. 
During the cooling, the electronic temperature dynamically changes, modifying also the coupling strength. 
With a calculated energy difference of the conduction band minimum at K and $\Sigma$ of 240~meV and optical phonon energies of 15-30~meV (at the corner of the BZ) \cite{Sahin2013}, photo-excited electrons emit approximately 10 phonons on average. 
The quantitative difference of single-state lifetime and measured \textbf{q}-dependent population dynamics is therefore in good agreement with expectations from a kinetic relaxation model. 

The population of zone-boundary phonons rises faster than the MSD and transiently overshoots compared to the thermal excited state reached after $\approx$50~ps. 
This indicates that the relaxation of photo-excited states in the conduction band is dominated by intervalley scattering, although the coupling is strongest for phonons facilitating intravalley scattering. 
We point out that hole-phonon scattering is neglected in our theory, for the following reason. 
The energy difference between the valence band maximum at $\Gamma$ and the top of the K valley valence band is approximately 50 meV \cite{Yuan2016}. 
Hole-phonon scattering hence contributes $\approx$17\% to the entire energy transferred from excited carriers to phonons. 
Subsequent to initial K-$\Gamma$ intervalley scattering, holes mainly relax within the $\Gamma$ valley through emission of small-\textbf{q} phonons. 
Hole-phonon scattering therefore will not significantly contribute to the phonon dynamics considered in this work.

To summarize, we have determined the coupling of photoexcited carriers in the conduction band of bulk \WSe\ to in-plane phonons by employing femtosecond electron diffraction and first principle DFT calculations. 
We find that the energy transfer from excited electrons to phonons occurs by carriers cooling within the $\Sigma$ valleys of the conduction band and the subsequent emission of phonons with a rate of few scatterings per picosecond. The electron-phonon scattering leads to a transient non-thermal phonon distribution at times below 50~ps, directly visualized by the experiment. Phonon-phonon and electron-phonon scattering eventually restore a thermal state at an elevated temperature. 

The experimental and theoretical momentum-dependence of the coupling are in good qualitative agreement and the quantitative differences are comprehensible. The momentum-dependent phonon dynamics can be understood by examining the available phase-space for electronic scattering, which is limited to intravalley and intervalley scattering between neighboring $\Sigma$ valleys. Our results indicate that intervalley scattering with zone-boundary phonons dominates the relaxation process.
We note that the electron-phonon scattering is expected to significantly change in monolayer \WSe, which is a direct-gap semiconductor \cite{Mak2010}. Although recent work suggests that dark excitons with electrons in the $\Sigma$-valley might exist in \WSe\ monolayers \cite{Selig2016}, the phase-space for electronic scattering is different and thus the momenta of emitted phonons. The agreement of the presented methods demonstrates their ability to resolve the momentum dependence of electron-phonon coupling, which will be crucial for applications employing electronic valley-coherence, e.g.~in TMDCs, for measuring microscopic electron-phonon and phonon-phonon coupling processes and for accurately predicting transport properties in a variety of materials.

This project has received funding from the Max Planck Society and from the European Research Council (ERC) under the European Union's Horizon 2020 research and innovation programme (grant agreement numbers ERC-2015-CoG-682843 and ERC-2015-AdG-694097).
R.B. thanks the Alexander von Humboldt foundation for financial support. 
H.H. acknowledges support from the People Programme (Marie Curie Actions) of the European Union's Seventh Framework Programme FP7-PEOPLE-2013-IEF project No.~622934. 
We acknowledges financial support from Grupos Consolidados (IT578-13) and Air Force Office of Scientific Research Award (Grant No.~FA2386-15-1-0006 AOARD 144088).

\bibliography{Waldecker_WSe2_diffraction}

\end{document}